\documentclass[%
 preprint,
 amsmath,amssymb,
 aps,
]{revtex4-1}
\usepackage{graphicx}
\usepackage{dcolumn}
\usepackage{bm}
\usepackage[pdftex,colorlinks]{hyperref}
\hypersetup{citecolor=blue}
\begin{document}
\title{Two schemes for characterization  and detection of  the squeezed light: Dynamical Casimir  effect and nonlinear materials} 
\author{H. Lotfipour}
\author{Z. Allameh}
\affiliation{Institute of Laser and Plasma of Shahid Beheshti University, Tehran, Iran\\}
 \author{R. Roknizadeh}
 \email{r.roknizadeh@gmail.com}
 \affiliation{
 Department of Physics, Quantum Optics Group\\University of Isfahan, Isfahan, Iran}
\author{H. Heydari}
\affiliation{
 Department of Physics, Stockholm University, Stockholm, Sweden\\}
 
\date{\today}

\begin{abstract}
The detection  and characterization  of a  non-classical-squeezed state of light, by using two different schemes, will be presented . In the  first one, in an one-dimensional cavity with moving mirror (non-stationary Casimir effect) in the principal mode,  we   study the {\it photon creation} rate for  two modes  (squeezed and coherent state) of driving   field.  Since the  cavity with moving mirror (similar to an  optomechanical system) can be considered  analogue  to a  Kerr-like medium, so that in the second scheme,  the probability  amplitude for  {\it multi-photon absorption} in a nonlinear (Kerr) medium  will be quantum mechanically calculated.  It is shown that because of presence of nonlinear effects, the responses of these two systems  to the squeezed versus coherent state are considerably  distinguishable. The drastic difference between the results of  these two  states of light can be viewed  as a proposal for detecting  of non-classical states.
\begin{description}

\item[PACS numbers]
42.50, 42.65
\end{description}
\end{abstract}

\pacs{Valid PACS appear here}
\maketitle

\section{introduction}
In recent years many experimental and theoretical progress in high-Q electromagnetic cavities have been made. Among them, there is a phenomenon that can be seen in the name of Non-Stationary Casimir Effect (NSCE) or Dynamical Casimir Effect (DCE) caused by changes of (vacuum) quantum states of fields due to a fast time variations of positions (or properties) of boundaries confining the fields \cite{Dodonov1,Dodonov2,Dodonov3}. The DCE was invented by Yablonovitch \cite{Yablonovitch} and Schwinger \cite{Schwinger} for the first time. The DCE has been observed in many systems such as: single mirror cavity \cite{Maia}, cavities with moving mirrors \cite{Dodonov1, Dodonov2, Dodonov3, Dodonov4}, circuits \cite{Wilson} and parametrically excited media \cite{Dezael}. Amplification of quantum (vacuum) fluctuations in these macroscopic systems is a common feature. In cavity with moving mirrors,  DCE concerns to  the reconstruction of the quantum state of a field, due to a time dependent geometrical configuration. Many phenomena \cite{Dodonov5} such as photon creation \cite{Dodonov4}, interaction between the field and detectors \cite{Castro,Dodonov4,Dodonov7}, generation of nonclassical (in particular squeezed) states \cite{Dodonov6}, etc. are observed in these cavities. Dodonov  {\it et al} have  investigated the photon creation in one-dimensional cavity in vacuum state (empty cavity) \cite{Dodonov4}.\\

Sousa {\it et al} \cite{Sousa} recently have shown that the nonlinear Kerr effect can be observed in cavities with DCE. In other words,  a  realistic description of cavity DCE  include   the Kerr nonlinearity.\\ 

In other side, optomechanics as a new field has some common features with DCE (which has been observed in cavity with moving mirror).  
Gong \emph{et al}, recently have showed that, the Kerr nonlinear optical effect can be observed in optomechanical system in vacuum. So the cavity with moving mirror can be supposed as a Kerr-like nonlinear medium \cite{Gong}.\\
 There are many standard methods for  generation  and detection of non-classical, in particular squeezed,  light. The squeezed light detection has been done in homodyne systems \cite{Shapiro,Teich, Breitenbach, Loundon}, which  is a phase-sensitive measurement of the optical field quadratures. The squeezed light interaction with atomic systems has been predicted in Refs. \cite{Dalton,Cessa}. While the most important advantage of cavity with moving wall (optomechanical system) is detection the quantum mechanical effects (such as squeezing) macroscopically.  
 
   Recently, some systems such as non-linear media \cite{02, 03, Zubairy} and optomechanical systems \cite{Fabre, Mancini} and resonators with moving wall \cite{Dodonov6} have been considered for generating squeezed light.  But, in nonlinear media with Optical Parametric Oscillator (OPO),  the squeezed parameters are affected by various  experimental features \cite{04,08,09}.
   
 Moreover nonlinear processes such as multiphoton absorption  can change the photon statistics of the incident light and generate a narrower photon distribution than poisson distribution \cite{01}. On the other hand, multiphoton absorption processes are very sensitive to the type of light  state and its fluctuations. The influence of nonclassical light  fluctuations particularly the squeezed states on the two photon absorption has been studied and its absorption rate compared with respect to the  phase and amplitude of the squeezed beam \cite{05}. The two photon interaction  of atoms and  squeezed light has been considered  in Refs.  \cite{Enaki, Schumaker}. The effect of the absorption process on the squeezed  light has been studied  in Ref. \cite{Enaki}. And the two photons quantum optics  has been investigated by \cite{Schumaker}. 
  In \cite{06}, as well, the multiphoton absorption rate is calculated by  quantum mechanical perturbation theory in a semi-classical approach. \\
 \indent Wilson  {\it et al} \cite{Wilson} in their  paper has experimentally shown the effect of DCE in superconducting circuit. They have  mentioned that this system operates as a parametric amplifier (nonlinear medium) to generate photons, and the mechanism of generating photons is very similar to a nonlinear medium.\\
 \indent In this contribution, as it has been predicted in Ref.  \cite{Wilson},  we are going to present two different, but somehow related, approaches for detecting the nonclassical-squeezed-state of light.\\
\indent Both approaches concern the nonlinear effects, which are sensitive to different modes of incident field. In the first part of the paper the interaction of squeezed state and a 1D cavity with moving mirror will be investigated. In the second one the interaction of squeezed light with nonlinear media is studied. The outcomes of these two schemes will be compared with the situation, when the incident field is in a coherent state. The drastic difference between two setups (squeezed and coherent states) can be used for characterizing the detection of the nonclassical light. \\
\indent In the first step,  we  will study the photon generation rate in  the one-dimensional cavity with moving wall (in the principal mode),  in which the  incident field is in coherent or  squeezed state.\\
In the second step, by quantizing the electromagnetic field, we  derive probability amplitude for three photon absorption in a nonlinear medium, by using quantum polarization \cite{07,010}.\\   
\indent In sec.(\ref{sec2}), we review the pioneer work of Dodonov and Klimov in Ref. \cite{Dodonov4}, to summarize the related results for calculating the photon generation rate in a cavity with moving boundaries. In sec.(\ref{sec3}) we derive photon generation rate in the cavity for two modes of light (squeezed and coherent light). The results for two different modes are also plotted in a few diagrams. The fundamental concepts of interaction of these two modes and nonlinear media  is explained in sec.(\ref{sec4}). In this section we obtain the rate of three photons absorption in a semi-classical approach And in sec.(\ref{sec5}), the problem of the last section is studied in full quantum mechanical approach. In sec.(\ref{sec6}) we obtain the rate of absorption for two kinds of light and compare them. Conclusions are presented in sec.(\ref{sec7}). \\
 \section{Review of Photon generation in cavity with moving boundary in vacuum state}\label{sec2}
In order the paper  be self contained, we review in this section the related results given in \cite{Dodonov4}. We consider a cavity having two infinite ideal plates, one being at rest at $ x=0 $, and the other one moving according to a time dependent function L(t) when $ t>0 $. We consider an electromagnetic field whose vector potential is directed along z axis. The field operator in \textit{Heisenberg representation} at $ t\leqslant0 $ (when both the plates were at rest) is  $ \hat{A}(x,t) $.
\begin{align}
\hat{A}_{in} = 2 \sum_{n} \dfrac{1}{\sqrt{n}}\sin\dfrac{n\pi x}{L_{0}}\hat{b}_{n}exp(-i\omega_{n}t)+c.c,\label{h1}
\end{align}
where $ \hat{b}_{n} $ is the annihilation photon operator and $ \omega_{n}=\dfrac{\pi n}{L_{0}} $.
The field operator can be written
\begin{align}
\hat{A}(x,t) = 2 \sum_{n} \dfrac{1}{\sqrt{n}}(\hat{b}_{n}\psi^{n}(x,t)+c.c.),\quad\quad n=1,2,\cdot\cdot\cdot \label{h2}
\end{align}
where
\begin{align}
\psi^{n}(x,t<0) =\sin\dfrac{n\pi x}{L_{0}}exp(-i\omega_{n}t).\label{h3}
\end{align}
We expend the function $ \psi^{n}(x,t) $ with respect to instantaneous basis\cite{Dodonov4}
\begin{align}
\psi^{n}(x,t>0) =\sum_{k} Q^{(n)}_{k}(t) \sqrt{\dfrac{L_{0}}{L(t)}}\sin\dfrac{\pi kx}{L(t)},\label{h4}
\end{align} 
where, for $ t>T $ (T is the time interval that the wall returns to its initial position $ L_{0} $)
\begin{align}
Q^{(n)}_{k}(t)=\xi^{(n)}_{k} e^{-i\omega_{k} t}+\eta^{(n)}_{k} e^{i\omega_{k} t},\label{h5}
\end{align}
$ \xi^{(n)}_{k} $ and $\eta^{(n)}_{k}$ are some constant coefficients. At $ t>T $ we introduce a new set of physical operators $ \hat{a}_{m} $ and $ \hat{a}^{\dagger}_{m} $. The Bogoliubov transformation relates the two sets of operators ($ \hat{a},\hat{a}^{\dagger} $ and $ \hat{b},\hat{b}^{\dagger} $) 
\begin{align}
\hat{a}_{m}=\sum_{n}(\hat{b}_{n}\alpha_{nm}+ \hat{b}^{\dagger}_{n}\beta^{\ast}_{nm}),\label{h6}
\end{align}
in which the coefficients are
\begin{align}
\alpha_{nm}=\sqrt{\dfrac{m}{n}}\xi^{(n)}_{m},\quad\quad\beta_{nm}=\sqrt{\dfrac{m}{n}}\eta_{m}^{(n)}.\label{h7}
\end{align}
We assume that the wall oscillates according to the following form,
\begin{align}
L(t)=L_{0}[1+\varepsilon\sin(2\omega_{1}t)],\quad\quad\quad\omega_{1}=\dfrac{\pi}{L_{0}}.\label{h8}
\end{align}
where $ \varepsilon\ll1 $.
We put the solution (6) in wave equation. After some simple algebraic manipulations we arrive to  the following equations in terms of coefficients $ \xi $ and $ \eta $:
\begin{align}
&\eta_{1}^{(3)}(\tau)=-\xi_{1}^{(1)}(\tau)-\dot{\eta}_{1}^{(1)}(\tau)\label{h9}
\end{align}
\begin{align}
&\xi_{1}^{(3)}(\tau)=-\eta_{1}^{(1)}(\tau)-\dot{\xi}_{1}^{(1)}(\tau)\label{h10}
\end{align}
\begin{align}
& n\xi_{1}^{(n+2)}(\tau)=n\xi_{1}^{(n-2)}(\tau)-\dot{\xi}_{1}^{(n)}(\tau) \quad ,\quad\quad\quad n\geq 3\label{h11}
\end{align}
\begin{align}
& n\eta_{1}^{(n+2)}(\tau)=n\eta_{1}^{(n-2)}(\tau)-\dot{\eta}_{1}^{(n)}(\tau)\quad ,\quad\quad\quad n\geq 3\label{h12}
\end{align}
where we have introduced new time scale:
\begin{align}
\tau=\dfrac{1}{2}\varepsilon\omega_{1}t.\label{h13} 
\end{align}
By using the Laplace transformation the solution of the above diffrential equations will be achieved,
and $\eta_{1}^{(1)}  $  and $ \xi_{1}^{(1)} $  are obtained  accordingly, 
\begin{align}
&\xi_{1}^{(1)}=\dfrac{2}{\pi}\dfrac{E(k)+\tilde{k}K(k)}{1+\tilde{k}}\label{h14}
\end{align}
\begin{align}
&\eta_{1}^{(1)}=-\dfrac{2}{\pi}\dfrac{E(k)-\tilde{k}K(k)}{1-\tilde{k}}\label{h15}
\end{align}
where $ K(k) $ and $ E(k) $ are complete elliptic integrals of the first and second kind:
\begin{align}
&K(k)=\int_{0}^{\pi/2} \dfrac{1}{\sqrt{1-k^{2}\sin^{2}(\alpha)}} \mathrm{d}\alpha \label{h16}
\end{align}
\begin{align}
&E(k)=\int_{0}^{\pi/2} \sqrt{1-k^{2}\sin^{2}(\alpha)} \mathrm{d}\alpha\label{h17}
\end{align}
And 
\begin{align}
k=\sqrt{1-e^{-8\tau}},\quad \tilde{k}=\sqrt{1-k^{2}}=e^{-4\tau}.\label{h18}
\end{align}
\section{inputing the squeezed  and coherent light}\label{sec3}
The amount of photons created in the \textit{m}th mode equals the average value of the operator 
$ \hat{a}^{\dagger}_{m}\hat{a}_{m} $, since just this opretator has a physical meaning at $ t>T $.
\subsection{Squeezed light}
We assume that the field in the cavity is initially in squeezed  coherent state $ |\alpha, \zeta\rangle_{b} $ where the subscript $ b $ indicates this state is defined with respect to the $\hat{b}_{n}$. The average of photon creation in mode $ m $ is given by,
\begin{align}
N_{m}=\ _{b}\langle\alpha, \zeta \vert\hat{a}^{\dagger}_{m}\hat{a}_{m} \vert\alpha,\zeta\rangle_{b}.\label{h19}
\end{align}
By using the Eq. \eqref{h6}, the above relation can be written in terms of  $ \hat{b} $ and $ \hat{b}^{\dagger} $ :
\begin{align}
N_{m} & =\ _{b}\langle\alpha, \zeta\mid \sum_{n^{\prime}}({\hat{b}}^{\dagger}_{n^{\prime}}\alpha^{\ast}_{n^{\prime}m} +{\hat{b}}_{n^{\prime}}\beta_{n^{\prime}m})\nonumber\\&\times\sum_{n}({\hat{b}}_{n}\alpha^{\ast}_{nm}+{\hat{b}}^{\dagger}_{n}\beta^{\ast}_{nm})\vert\alpha,\zeta \rangle_{b}\nonumber\\
&=\ _{b}\langle\alpha , \zeta\vert \sum_{n^{\prime}}\sum_{n}
(\alpha^{\ast}_{n^{\prime}m}\alpha_{nm}\hat{b}^{\dagger}_{n^{\prime}}\hat{b}_{n}
+\alpha^{\ast}_{n^{\prime}m}\beta^{\ast}_{nm}\hat{b}^{\dagger}_{n^{\prime}}\hat{b}^{\dagger}_{n}
\nonumber\\ &+\alpha_{nm}\beta_{n^{\prime}m}\hat{b}_{n^{\prime}}\hat{b}_{n}
+\beta_{n^{\prime}m}\beta^{\ast}_{nm}\hat{b}_{n^{\prime}}\hat{b}^{\dagger}_{n})\vert\alpha, \zeta\rangle_{b}.\label{h20}
\end{align}
This expectation value can be calculated  by using the squeezing operator according to the following relations \cite{015}, 
\begin{align}
&\mid\alpha ,\zeta\rangle =\hat{s}(\zeta)\mid\alpha\rangle ,\quad\quad &(a)\nonumber\\
&\hat{s}^{\dagger}(\zeta)\hat{b}_{n}\hat{s}(\zeta)=\mu\hat{b}_{n}-\nu\hat{b}^{\dagger}_{n},\quad\quad &(b)\nonumber\\
&\hat{s}^{\dagger}(\zeta)\hat{b}^{\dagger}_{n}\hat{s}(\zeta)=\mu\hat{b}^{\dagger}_{n}-\nu^{\ast}\hat{b}_{n},\quad\quad &(c)\nonumber\\
&\hat{b}_{n}\mid\alpha\rangle=\alpha\mid\alpha\rangle .\quad\quad &(d)\label{h21}
\end{align}
where $\hat {s}(\zeta)  $ is squeezing operator and $ \mu=\cosh\vert\zeta\vert ,\nu=e^{i\varphi}\sinh\vert\zeta\vert$, and $\zeta,  \varphi $ are squeezing parameter and its phase respectively.
After some calculation Eq. \eqref{h20} gives rise to,
\begin{align}
N_{m}&=\nonumber\\&\sum_{n}\lbrace\vert\alpha_{nm}\vert^{2}[\mu^{2}\vert\alpha\vert^{2}-2\mu \text{Re}(\nu^{\ast}\alpha^{2})+\vert\nu\vert^{2}\vert\alpha\vert ^{2}
+\vert\nu\vert^{2}]\nonumber\\&+2\text{Re}(\beta_{nm}\alpha_{nm})
\text{Re}[\mu^{2}\alpha^{2}-2\mu\nu\vert\alpha\vert^{2}-\mu\nu+\nu^{2}\alpha^{{\ast}^{2}}]\nonumber\\&
+\vert\beta_{nm}\vert^{2}
[\mu^{2}\vert\alpha\vert^{2}-\nu^{2}\vert\alpha\vert^{2}+\mu^{2}-2\mu\text{Re}(\nu^{\ast}\alpha^{2})]\rbrace.\label{h23}
\end{align}
The photon generation rate   in principal cavity mode [with $ m=1 $] is given according to,
\begin{align}
\dfrac{dN}{d\tau} &=\nonumber\\&\sum_{n}\lbrace\dfrac{2}{n}\vert\xi_{1}^{(n)}\dot{\xi}_{1}^{(n)}\vert(\mu^{2}\vert\alpha\vert^{2}\nonumber\\&-2\mu \text{Re}(\nu^{\ast}\alpha^{2})+\vert\nu\vert^{2}\vert\alpha\vert ^{2}
+\vert\nu\vert^{2})\nonumber\\&+\dfrac{2}{n}\text{Re}(\dot{\xi}_{1}^{(n)}\eta_{1}^{(n)}+\xi_{1}^{(n)}\dot{\eta}_{1}^{(n)})
\text{Re}(\mu^{2}\alpha^{2}-2\mu\nu\vert\alpha\vert^{2}\nonumber\\&-\mu\nu +\nu^{2}\alpha^{{\ast}^{2}})
+\dfrac{2}{n}\vert\eta_{1}^{(n)}\dot{\eta}_{1}^{(n)}\vert(\mu^{2}\vert\alpha\vert^{2}-2\mu \text{Re}(\nu^{\ast}\alpha^{2})\nonumber\\&-\nu^{2}\vert\alpha\vert^{2}+\mu^{2})\rbrace.\label{h24}
\end{align}
where we have used of Eq. \eqref{h7}. The sums in Eq. \eqref{h24} can be simplified 
\begin{align}
\sum_{n=1}^{\infty}\dfrac{2}{n}\eta_{1}^{(n)}\dot{\eta}_{1}^{(n)}
=-2\eta_{1}^{(1)}\xi_{1}^{(1)},\label{h25}
\end{align}
\begin{align}
\sum_{n=1}^{\infty}\dfrac{2}{n}\xi_{1}^{(n)}\dot{\xi}_{1}^{(n)}
=-2\eta_{1}^{(1)}\xi_{1}^{(1)},\label{h26}
\end{align}
\begin{align}
\sum_{n=1}^{\infty}\dfrac{1}{n}(\dot{\xi}_{1}^{(n)}\eta_{1}^{(n)}+\xi_{1}^{(n)}\dot{\eta}_{1}^{(n)})
=-\eta_{1}^{{(1)}^{2}}-\xi_{1}^{{(1)}^{2}}.\label{h27}
\end{align}
Therefore  we obtain  the final form of the generation rate,
\begin{align}
\dfrac{dN}{d\tau}&=\nonumber\\
&(-2\eta_{1}^{(1)}\xi_{1}^{(1)})(\mu^{2}\vert\alpha\vert^{2}-2\mu \text{Re}(\nu^{\ast}\alpha^{2})+\vert\nu\vert^{2}\vert\alpha\vert ^{2}
+\vert\nu\vert^{2})\nonumber\\&+2\text{Re}(-\eta_{1}^{{(1)}^{2}}-\xi_{1}^{{(1)}^{2}})
\text{Re}(\mu^{2}\alpha^{2}-2\mu\nu\vert\alpha\vert^{2}-\mu\nu +\nu^{2}\alpha^{{\ast}^{2}})\nonumber\\&
+(-2\eta_{1}^{(1)}\xi_{1}^{(1)})(\vert\alpha\vert^{2}(\mu^{2}+\vert\nu\vert^{2})-2\mu \text{Re}(\nu^{\ast}\alpha^{2})+\mu^{2}).\label{h28}
\end{align}

\subsection{Coherent light}
\begin{figure}[ht]
\begin{center}
\includegraphics[scale=0.27]{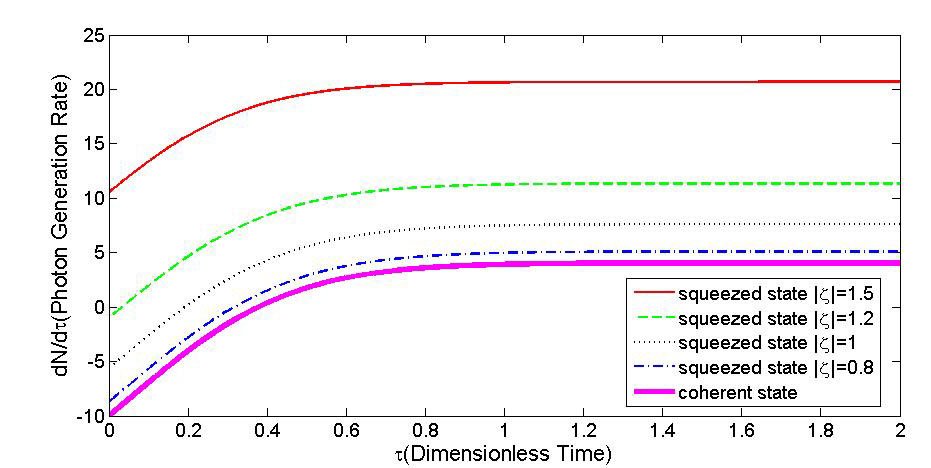}
\end{center}
\caption{Photon generation rate versus the dimensionless time $\tau$ for squeezed light with various squeezed parameters ($\vert\zeta\vert$ = 0.8,  $\vert\zeta\vert$ = 1,  $\vert\zeta\vert$ = 1.2  and $\vert\zeta\vert$ = 1.5 , $ \varphi_{\zeta}=0 $.) and coherent light ($ \vert\alpha\vert^{2}=7 $) \cite{Mancini}.}\label{plot1}
\end{figure}

Now we  apply a driving field in  coherent state, and follow the same procedure as given in the above subsection in order to calculate the photon generation rate.The field is considered to be in the coherent state $ |\alpha\rangle_{b} $, where the subscript  $b$ indicates that these states are eigenvectors of the annihilation operator $\hat{b}_{n}$. The photon number created in the \textit{m}th mode  is given by $N_{m} $,
\begin{align}
N_{m}=\ _{b}\langle\alpha \mid \hat{a}^{{\dagger}_{m}}\hat{a}_{m} \mid\alpha \rangle_{b},\label{h29}
\end{align}
and as we have just now noted,
\begin{align}
\hat{b}\mid\alpha\rangle_{b}=\alpha\mid\alpha\rangle_{b},\label{h30}
\end{align}
where the   expansion in terms of Fock states is well known,
\begin{align}
|\alpha\rangle_{b}=\exp(\dfrac{-\mid\alpha\mid^{2}}{2})\sum_{p=0}^{\infty}\dfrac{\alpha^{p}}{\sqrt{p!}}\mid p\rangle .\label{h31}
\end{align}
The action of $ \hat{a}_{m} $ on this state will be calculated by using the transformation relation in Eq. \eqref{h6}, 
\begin{align}
\hat{a}_{m} |\alpha\rangle_{b}&=\nonumber\\&\exp(\dfrac{-\mid\alpha\mid^{2}}{2})\sum_{n}\sum_{p=0}^{\infty}\dfrac{\alpha^{p}}{\sqrt{p!}}(\alpha_{nm}\delta_{np}\sqrt{p}\mid p-1\rangle\nonumber\\&+\beta^{\ast}_{nm}\delta_{np}\sqrt{p+1}\mid p+1\rangle)
\nonumber\\&=\sum_{n}\exp(\dfrac{-\mid\alpha\mid^{2}}{2})\dfrac{\alpha^{n}}{\sqrt{n!}}(\alpha_{nm}\sqrt{n}\mid n-1\rangle\nonumber\\&
+\beta^{\ast}_{nm}\sqrt{n+1}\mid n+1\rangle).\label{h32}
\end{align}
By applying the same procedure we can get the action of $ \hat{a}_{m} $ on $ |\alpha\rangle_{b} $, and accordingly,
\begin{align}
N_{m}=\sum_{n=0}^{\infty}\exp(-\mid\alpha\mid^{2})\dfrac{\alpha^{2n}}{n!}(n\alpha_{nm}^{2}
+\beta_{nm}^{{\ast}^{2}}(n+1)),\label{h33}
\end{align}
and by replacing $ \alpha_{nm} $ and $ \beta_{nm} $ from Eq. \eqref{h7},
\begin{align}
N_{m}=\sum_{n=0}^{\infty}\exp(-\mid\alpha\mid^{2})\dfrac{\alpha^{2n}}{n!}
(m{\xi}_{m}^{{(n)}^{2}}+\dfrac{n+1}{n}m\eta_{m}^{{\ast}^{{(n)}^{2}}}).\label{h34}
\end{align}
The photon generation rate in principal cavity mode [with $ m=1 $] is
\begin{align}
\dfrac{dN}{d\tau}&
=\exp(-\mid\alpha\mid^{2})\sum_{n=0}^{\infty}\alpha^{2n}\dfrac{2\xi^{(n)}_{1}\dot{\xi}_{1}^{(n)}}{n!}
\nonumber\\&
+\exp(-\mid\alpha\mid^{2})\sum_{n=0}^{\infty}\alpha^{2n}\dfrac{2}{n!}\dfrac{n+1}{n}\eta_{1}^{{\ast}^{(n)}}\dot{\eta}_{1}^{{\ast}^{(n)}}.\label{h35}
\end{align}
After some calculation and using of Eqs. \eqref{h27}-\eqref{h30}, the Eq. \eqref{h35} can be rewrittn, 
\begin{align}
\dfrac{dN}{d\tau}&=-4\eta_{1}^{(1)}\xi_{1}^{(1)}\mid\alpha\mid^{2}-2\eta_{1}^{(1)}\xi_{1}^{(1)}\nonumber\\&+2\text{Re}(-\eta_{1}^{{(1)}^{2}}
-\xi_{1}^{{(1)}^{2}})\text{Re}(\alpha^{2}).\label{h36}
\end{align}
Now, we can compare the photon generation rate for the cavity mode in coherent and squeezed state, according to the Eq. \eqref{h28} and Eq. \eqref{h36}, where the explicit forms of $ \eta_{1}^{(1)} $ and $ \xi_{1}^{(1)} $, as given in relations \eqref{h14}-\eqref{h18}, are to be used. It should be considered that the intensity of coherent and squeezed light  is the same and equals to  7 (The coherent state amplitude ,$ \alpha $, is a complex number).\\
\indent Their plots are shown in Fig. \ref{plot1}. 

The drastic difference between different squeezed lights and coherent light in photon generation rate may be easily seen. The effect of squeezing parameter has been shown in  the same figure. The photon generation rate increases with increasing of squeezing parameter. This point is considerable that as usual, the squeezing phenomenon occurs in special region, so we restrict ourself in the related  region \cite{015}.

\begin{figure}[ht]
\begin{center}
\includegraphics[scale=0.26]{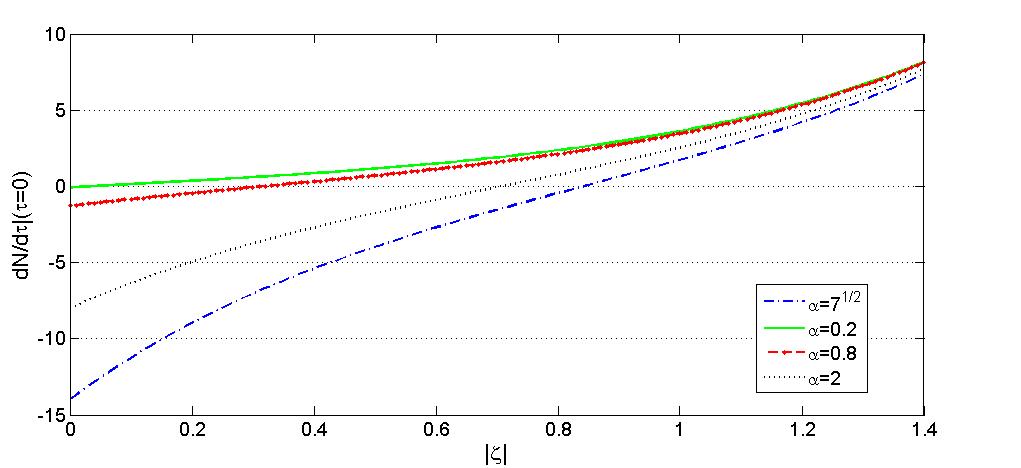}
\end{center}
\caption{Plot of $ \dfrac{dN}{d\tau}\vert_{\tau=0} $ as a function of the squeezing parameter $ \vert\zeta\vert $ for different values of real coherent state amplitude \quad $ \alpha= 0.2, 0.8, 2.0, \sqrt{7}$.} \label{Fig1}
\end{figure}

\begin{figure}[ht]
\begin{center}
\includegraphics[scale=0.26]{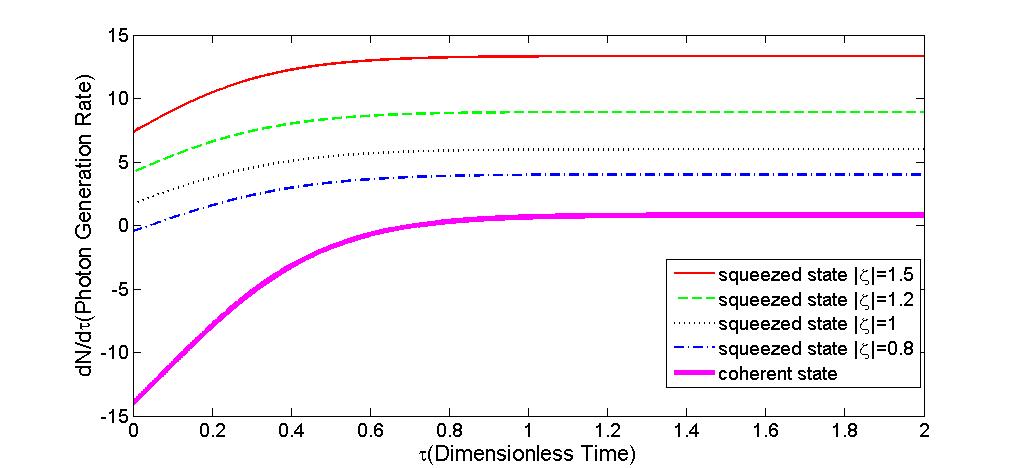}
\end{center}
\caption{Photon generation rate for squeezed and coherent states for real value of $ \alpha $.} \label{Fig2}
\end{figure}

\begin{figure}[ht]
\begin{center}
\includegraphics[scale=0.27]{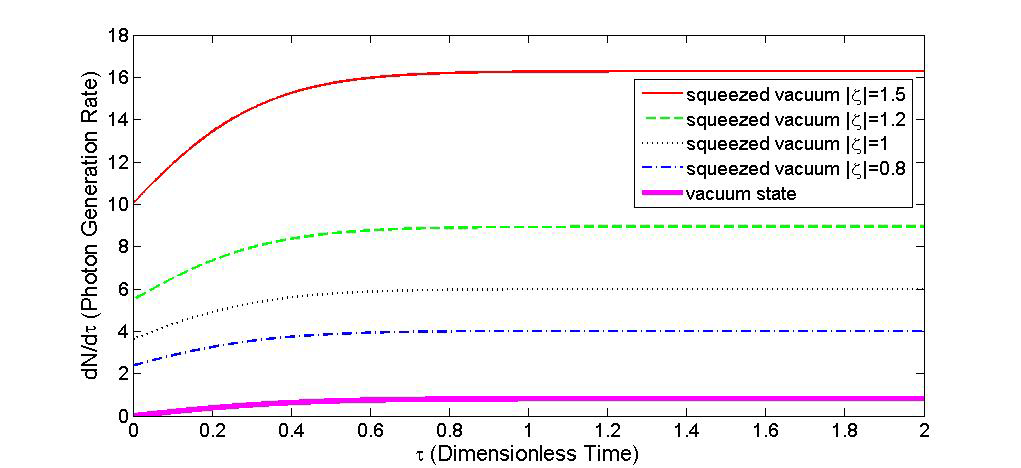}
\end{center}
\caption{Photon generation rate versus the dimensionless time $ \tau $ for vacuum state and squeezed vacuum state with various squeezed parameters ($\vert\zeta\vert$ = 0.8,  $\vert\zeta\vert$ = 1,  $\vert\zeta\vert$ = 1.2  and $\vert\zeta\vert$ = 1.5 , $ \varphi_{\zeta}=0 $.)} \label{plot2}
\end{figure}

It is required to investigate Eq. \eqref{h28} in some limits. In the limit $ \tau\rightarrow0  $ the variables $ k\rightarrow0 $ , \quad $\tilde{k}\rightarrow1$, \quad $ \lbrace E(k),K(k)\rbrace\rightarrow\dfrac{\pi}{2}$, \; $\eta_{1}^{(1)}\rightarrow0 $ , and $ \xi^{(1)}_{1}\rightarrow1 $ . The photon generation rate in the limit $\tau\rightarrow0$ for squeezed state $ (\alpha\neq0) $ has the form,
\begin{align}
\dfrac{dN}{d\tau}&=-2\text{Re}(\mu^{2}\alpha^{2}-2\mu\nu\vert\alpha\vert^{2}-\mu\nu+\nu^{2}\alpha^{{\ast}^{2}})\quad for \quad \tau\rightarrow0,\label{h37}
\end{align}
which , for a fixed real value of $ \alpha  $, is a monotonic function of $ \vert\zeta\vert $ (Fig \ref{Fig1}).\\
The photon generation rate (for squeezed and coherent states) for real value of $ \alpha $ is plotted in Fig. \ref{Fig2} which concides with Eq. \eqref{h37} and Fig. \ref{Fig1}.\\
For completeness the the amount of the photon creation, when the driving field  are in vacuum  $|0\rangle_b$ and squeezed vacuum states( with different squeezing parameters), is plotted in Fig. \ref{plot2}. The relation for vacuum state in this case coincides with the Eq. (6.1) in \cite{Dodonov4}. According to the Eq. \eqref{h37}, $ \dfrac{dN}{d\tau}\vert_{\tau=0} $ is equal to $ 2\mu\nu $ which is a positive value. The initial values of photon generation rate in Fig. \ref{plot2}  are positive.  
 
\section{The photon absorption rate in nonlinear media}\label{sec4}

Gong in his paper \cite{Gong} derived a Kerr-medium-like Hamiltonian in the vacuum induced by the radiation pressure for optomechanical systems which shows the underlying nonlinear mechanism more clearly than other approaches. In this mechanism the third-order susceptibility can be obtained as in usual Kerr media.\\ 
So we can simulate cavity with moving mirror as a  Kerr-medium-like with non-linear susceptibility. With regard to the relationship between the cavity with moving mirror and nonlinearity it is needed that the effect of light on nonlinear media is explored. Therefore in the present and next sections we investigate the effects of squeezed and coherent light on a nonlinear media with third-order susceptibility.\\ 
\indent  The multiphoton absorption rates are, first,  studied in semi-classical approach \cite{06}. At this scheme, the electric field is considered classical and the atomic wave function is the solution of the time-dependent Schr\"{o}dinger  
equation,\\
\begin{equation}
i\hbar\dfrac{\partial\psi(r,t)}{\partial t}=\hat{H}\psi(r,t)\label{1},
\end{equation} 
where 
\begin{equation}
\hat{H}=\hat{H_{0}}+\hat{V},\label{1a}
\end{equation}
here $ H $ is the Hamiltonian operator of the system. $ \hat{H_{0}} $ is the Hamiltonian for a free system and $ \hat{V} $ is the interaction energy in the form, 
\begin{equation}
\hat{V}=-\hat{d} {E}(t),\label{2}
\end{equation}
where $ \hat{d}=e\hat{r} $ is the electric dipole moment, $ \hat{r} $ and $ -e $ is the displacement operator and the charge of electron, respectively. To find the general solution for the system, the standard ansatz is made: \\
\begin{equation}\label{3}
\psi(r,t)=\sum_{l}h_{l}(t)u_{l}(r)e^{-i\omega_{l} t},
\end{equation}
here $ h_{l}(t) $ gives the probability amplitude and $ u_{l}(r) $  satisfy the eigenvalue equation, when the interaction is not present, \\ 
\begin{equation}
\hat{H_{0}}u_{l}(r)=E_{l}u_{l}(r)\qquad\qquad E_{l}=\hbar \omega_{l}.\label{4}
\end{equation}
The eigenstates constitute a complete orthonormal  set satisfying the condition  \\
\begin{equation*}
\int u_{l}(r)u_{m}^{*}(r)\text{d}r=\delta_{lm}.
\end{equation*}
Therefore the Schr\"{o}dinger equation is written in the following form,
\begin{equation}
i\hbar \dfrac{d h_{m}}{d t}=\sum_{l}h_{l}(t)V_{ml}(r)e^{-i\omega_{lm} t},\label{5}
\end{equation}
 where $ \omega_{lm}=\omega_{l}-\omega_{m} $ and $ V_{ml} $ are the matrix elements of the interaction  $ \hat{V} $, 
\begin{equation}
V_{ml}=\int u^{*}_{m}(r)\hat{V}u_{l}(r) d^{3}r\label{6}.
\end{equation} 
Eq. \eqref{5} can be solved by using   the perturbation theory. The Hamiltonian Eq. \eqref{1a} is replaced by
\begin{equation}
\hat{H}=\hat{H_{0}}+\lambda\hat{V},\label{6a}
\end{equation}
 here $ \lambda $  is expansion parameter. Also, the $h_l(t)$ in Eq.\eqref{5} and $\psi(r,t) $ in Eq.\eqref{3} can be expanded in powers of $ \lambda $:
\begin{align}
&h_{l}(t)=h_{l}^{(0)}(t)+\lambda h_{l}^{(1)}(t)+\lambda^{2}h_{l}^{(2)}(t)+...,\nonumber\\
&\psi(r,t)=\psi^{(0)}(r,t)(t)+\lambda \psi^{(1)}(r,t)(t)+\lambda^{2}\psi^{(2)}(r,t)(t)+...,\label{7}
\end{align} 
here $ h_{l}^{(0)} $ corresponds to the ground state in the absence of the applied field and we set it equal to $ 1 $. So $ \psi^{(0)}(r,t) $ can be considered as,
\begin{equation}
 \psi^{(0)}(r,t)= u_{g}(r)e^{-i\omega_{g}t},\qquad l=g\label{7a}
\end{equation}
 By equating powers of $ \lambda $ on each side in Eq. \eqref{5} the set of equations for different orders of $ h_{l}(t) $ and $ h_{m}(t) $ are obtained.\\
\begin{align}
\dfrac{dh_{m}^{N}(t)}{dt}=&(i\hbar)^{-1}\sum_{l}h_{l}^{(N-1)}(t)V_{ml}(r)e^{-i\omega_{lm} t},\nonumber\\
& N=1,2,3,...\label{8}
\end{align}   
 
For simplicity the electric field is considered as a monochromatic wave of the form:\\
\begin{equation}\label{9}
E(t)=E e^{-i\omega t}+c.c.
\end{equation} 
The  $ N- $photon absorption rate for the atom in state $ m $ at time $ t $ is given by
\begin{equation}\label{10}
R_{mg}^{(N)}=\dfrac{\text{dP}_{m}^{(N)}(t)}{\text{dt}},
\end{equation} 
where $P_{m}^{(N)}(t)=\vert h_{m}^{(N)}(t)\vert^{2}  $ is $ N- $photon absorption probability. Now let us to consider a three level atom with $ (g,m,n,l) $ states( see Fig.\ref{pz4}).  It is straightforward to calculate the transition rate for third photon absorption in semi-classical approach\cite{06}:
\begin{align}\label{11}
R^{3}_{l}=\vert\sum_{mn}\dfrac{d^{gm}d^{mn}d^{nl}E^{3}}{\hbar^{3}(\omega^{mg}-\omega)(\omega^{ng}-2\omega)}\vert^{2}2\pi\rho_{f}(\omega^{lg}-3\omega).
\end{align}
Here $ \rho_{f}(\omega) $ denotes the density of final state, as a distribution on the frequency range. $ d^{gm} $ is the electric dipole transition moment between $ g $(ground state) and  state $m$. Also the polarization for this system  is derived  by applying  the third order perturbation,
\begin{align}
\langle p^{(3)}\rangle =&\langle \psi^{(0)}\vert \hat{d}\vert \psi^{(3)}\rangle +\langle \psi^{(1)}\vert \hat{d}\vert \psi^{(2)}\rangle +\nonumber\\
&\langle \psi^{(2)}\vert \hat{d}\vert \psi^{(1)}\rangle +\langle \psi^{(3)}\vert \hat{d}\vert \psi^{(0)}\rangle \label{12}.
\end{align}
 and by using Eqs. \eqref{3}, \eqref{6},\eqref{7} and \eqref{7a} one gets
\begin{align}\label{13}
\langle p^{(3)}\rangle =&\sum_{l}h_{l}^{(3)}(t) d^{gl}(r)e^{-i\omega_{lg} t}+\nonumber\\
&\sum_{nm}h_{n}^{(2)}(t)h_{m}^{*(1)}(t) d^{mn}(r)e^{-i\omega_{nm} t}+c.c. 
\end{align}
 For simplicity it is assumed that virtual levels $ (l,m,n) $ are nearly coincident with real atomic levels. On the other hand, by using second quantization, the average of polarization operator can be obtained. At this scheme, by introducing Dirac notation for $ u_{\beta}(r) $ as $ \vert \beta\rangle $, the orthogonality and completeness relation take the forms,
\begin{align}\label{14}
&\langle \beta\vert \beta^{'}\rangle=\delta_{\beta\beta^{'}}\nonumber\\
&\sum_{\beta}\vert\beta\rangle\langle\beta\vert\equiv 1.
\end{align}
The following identification is well understood, 
\begin{equation}
\vert\beta\rangle\langle\beta^{'}\vert\equiv \hat{c}^{\dagger}_{\beta}\hat{c}_{\beta^{'}}\label{15},
\end{equation}
here operator $\hat{c}_{\beta^{'}} (\hat{c}^{\dagger}_{\beta}) $ annihilates (creates) electron at the state $ \vert\beta^{'}\rangle(\vert\beta\rangle) $.
So that we obtain the  form of the polarization operator $ \hat{p}=e\hat{r} $,
\begin{equation}
\hat{p}=\sum_{\beta\beta^{'}}\vert\beta\rangle\langle\beta\vert e\hat{r}\vert\beta^{'}\rangle\langle\beta^{'}\vert =\sum_{\beta\beta^{'}}d^{\beta\beta^{'}}\langle\hat{c}^{\dagger}_{\beta}\hat{c}_{\beta^{'}}\rangle \label{16}.
\end{equation}
This form of polarization is general and includes all possible situations that may be occured, so we choose the part that is related to the third order of polarization,
\begin{equation}
\langle p^{(3)}\rangle =d^{gl}\langle\hat{c}^{\dagger}_{g}\hat{c}_{l}\rangle +d^{mn}\langle\hat{c}^{\dagger}_{m}\hat{c}_{n}\rangle +c.c\label{17}.
\end{equation}
Comparing  the  Eq. \eqref{17} and Eq. \eqref{13}, we find that $ h_{l}^{3}(t)e^{i\omega_{gl}t} $ in semi-classical approach is equivalent to $ \langle \hat{c}_{g}^{\dagger}\hat{c}_{l}\rangle $ in quantum approach,
\begin{equation}
\vert h_{l}^{(3)}(t)\vert^{2}\equiv \vert  \langle \hat{c}_{g}^{\dagger}\hat{c}_{l}\rangle\vert^{2}\label{18}.
\end{equation} 
Now, we need to evaluate the dynamics of microscopic polarization operator $ \langle\hat{p}^{lg}\rangle = \langle \hat{c}_{l}^{\dagger}\hat{c}_{g}\rangle $.
\section{quantum approach}\label{sec5}
The system under study is schematically  shown in Fig. \ref{pz4}.
\begin{figure}[ht] 
\begin{center}
\includegraphics[scale=0.5]{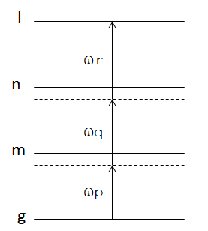}\label{pz4}
\end{center}
\caption{Schematic representation of the third-order polarization.}\label{pz4}
\end{figure}\\
The total Hamiltonian for this system, includes two main parts,  
\begin{equation}
H_{tot}=H_{0 }+H_{D}\label{19},
\end{equation}
here $ H_{0} $  includes the non-interacting parts of carrier and  photons system. The dipole interaction between light and atom is given by $ H_{D} $. The explicit form of Hamiltonian in second quantization scheme as is given in \cite{013,014,015}, 
\begin{align}
H_{0}=&\varepsilon^{g}\hat{c}^{\dagger}_{g}\hat{c}_{g}+\varepsilon^{m}\hat{c}^{\dagger}_{m}\hat{c}_{m}+\varepsilon^{n}\hat{c}^{\dagger}_{n}\hat{c}_{n}+\varepsilon^{l}\hat{c}^{\dagger}_{l}\hat{c}_{l}\nonumber\\
+&\hbar\omega_{p}(\hat{b}^{\dagger}_{p}\hat{b}_{p}+\dfrac{1}{2})+\hbar\omega_{q}(\hat{b}^{\dagger}_{q}\hat{b}_{q}+\dfrac{1}{2})\nonumber\\
+&\hbar\omega_{r}(\hat{b}^{\dagger}_{r}\hat{b}_{r}+\dfrac{1}{2})\nonumber ,\\
H_{D}=&-\lbrace(i\hbar \mathcal{F}_{p}d^{mg}\hat{c}^{\dagger}_{m}\hat{c}_{g}\hat{b}_{p}-i\hbar \mathcal{F}_{p}^{*}d^{gm}\hat{c}^{\dagger}_{g}\hat{c}_{m}\hat{b}_{p}^{\dagger})\nonumber\\ 
+&(i\hbar \mathcal{F}_{q}d^{nm}\hat{c}^{\dagger}_{n}\hat{c}_{m}\hat{b}_{q}-i\hbar \mathcal{F}_{q}^{*}d^{mn}\hat{c}^{\dagger}_{m}\hat{c}_{n}\hat{b}_{q}^{\dagger})\nonumber\\
+&(i\hbar\mathcal{F}_{r}d^{ln}\hat{c}^{\dagger}_{l}\hat{c}_{n}\hat{b}_{r}-i\hbar \mathcal{F}_{r}^{*}d^{nl}\hat{c}^{\dagger}_{n}\hat{c}_{l}\hat{b}_{r}^{\dagger})\rbrace \label{20},
\end{align}
where $ \mathcal{F}_{p} $ is the interaction strength between carriers  and photon with mode indx $ p $.  $ \hat{b}_{p} (\hat{b}^{\dagger}_{p}) $ is the photon annihilation (creation) operator. By using the Heisenberg equation of motion the dynamics of the expectation value of polarization is given by
\begin{align}
i\hbar\dfrac{\partial}{\partial t}\langle \hat{c}_{l}^{\dagger}\hat{c}_{g}\rangle &=\langle[\hat{c}_{l}^{\dagger}\hat{c}_{g},H_{tot}]\rangle ,\nonumber\\
i\hbar\dfrac{\partial}{\partial t}\langle \hat{c}_{l}^{\dagger}\hat{c}_{g}\rangle &=(\varepsilon^{g}-\varepsilon^{l})\langle \hat{c}_{l}^{\dagger}\hat{c}_{g}\rangle -\lbrace -i\hbar \mathcal{F}_{p}^{*}d^{gm}\langle\hat{c}^{\dagger}_{l}\hat{c}_{m}\hat{b}_{p}^{\dagger}\rangle\nonumber\\ 
&+i\hbar \mathcal{F}_{r}^{*}d^{nl}\langle\hat{c}^{\dagger}_{n}\hat{c}_{g}\hat{b}_{r}^{\dagger}\rangle)\rbrace \label{21}.
\end{align}
In order to solve this quantum excitation, we have to evaluate the dynamics of the photon-assisted polarization. Applying the Heisenberg equation of motion for $ \langle\hat{c}^{\dagger}_{l}\hat{c}_{m}\hat{b}_{p}^{\dagger}\rangle $ leades to:
\begin{align}
i\hbar\dfrac{\partial}{\partial t}\langle\hat{c}^{\dagger}_{l}\hat{c}_{m}\hat{b}_{p}^{\dagger}\rangle &=(-\varepsilon^{l}+\varepsilon^{m}-\hbar\omega_{p})\langle\hat{c}^{\dagger}_{l}\hat{c}_{m}
\hat{b}_{p}^{\dagger}\rangle \nonumber\\
& + i\hbar \mathcal{F}_{q}^{*}d^{{nm}^{*}}\langle \hat{c}^{\dagger}_{l}\hat{c}_{n}\hat{b}^{\dagger}_{p}\hat{b}^{\dagger}_{q}\rangle  \nonumber\\
&- i\hbar \mathcal{F}_{p}d^{mg}\langle \hat{c}^{\dagger}_{l}\hat{c}_{g}\hat{b}_{p}\hat{b}^{\dagger}_{p}\rangle \nonumber\\
&- i\hbar \mathcal{F}_{r}^{*}d^{{ln}^{*}}\langle \hat{c}^{\dagger}_{n}\hat{c}_{m}\hat{b}^{\dagger}_{r}\hat{b}^{\dagger}_{p}\rangle \nonumber\\
&-i\hbar \mathcal{F}_{p}d^{mg}\langle\hat{c}^{\dagger}_{l}\hat{c}_{m}\hat{c}^{\dagger}_{m}\hat{c}_{g}\rangle .\label{22}
\end{align}
and for $ \langle\hat{c}^{\dagger}_{n}\hat{c}_{g}\hat{b}_{r}^{\dagger}\rangle $ one obtains:
\begin{align} 
i\hbar\dfrac{\partial}{\partial t}\langle\hat{c}^{\dagger}_{n}\hat{c}_{g}\hat{b}_{r}^{\dagger}\rangle &=(-\varepsilon^{n}+\varepsilon^{g}-\hbar\omega_{r})\langle\hat{c}^{\dagger}_{n}\hat{c}_{g}
\hat{b}_{r}^{\dagger}\rangle\nonumber\\
& + i\hbar \mathcal{F}_{p}^{*}d^{{mg}^{*}}\langle \hat{c}^{\dagger}_{n}\hat{c}_{m}\hat{b}^{\dagger}_{r}\hat{b}^{\dagger}_{p}\rangle  \nonumber\\
&+ i\hbar \mathcal{F}_{r}d^{ln}\langle \hat{c}^{\dagger}_{l}\hat{c}_{g}\hat{b}_{r}\hat{b}^{\dagger}_{r}\rangle\nonumber\\
&- i\hbar \mathcal{F}_{q}^{*}d^{{nm}^{*}}\langle \hat{c}^{\dagger}_{m}\hat{c}_{g}\hat{b}^{\dagger}_{q}\hat{b}^{\dagger}_{r}\rangle\nonumber\\
&+i\hbar \mathcal{F}_{r}d^{ln}\langle\hat{c}^{\dagger}_{l}\hat{c}_{n}\hat{c}^{\dagger}_{n}\hat{c}_{g}\rangle \label{23}.
\end{align}
Eq. \eqref{22} and Eq. \eqref{23} have the same intuitive interpretation. In Eq. \eqref{22}, the photon-assisted polarization is generated either via spontaneous emission or via absorbing a new photon, including the $ \langle \hat{c}^{\dagger}_{l}\hat{c}_{n}\hat{b}^{\dagger}_{p}\hat{b}^{\dagger}_{q}\rangle $ and $ \langle \hat{c}^{\dagger}_{l}\hat{c}_{g}\hat{b}_{p}\hat{b}^{\dagger}_{p}\rangle $ terms, respectively. It may be also transform to a new term, via spontaneous emission by $ \langle \hat{c}^{\dagger}_{n}\hat{c}_{m}\hat{b}^{\dagger}_{p}\hat{b}^{\dagger}_{r}\rangle $ term. The latter term contains the correlated carriers. For our purpose, it is sufficient to consider only the terms which produce photon-assisted by spontaneous emission. So we consider the other terms as decay terms. The dynamics of these terms follow from Heisenberg equation of motion,
\begin{align}
i\hbar\dfrac{\partial}{\partial t}\langle \hat{c}^{\dagger}_{l}\hat{c}_{n}\hat{b}^{\dagger}_{p}\hat{b}^{\dagger}_{q}\rangle =&(-\varepsilon^{l}+\varepsilon^{n}-\hbar\omega_{p}-\hbar\omega_{q})\langle \hat{c}^{\dagger}_{l}\hat{c}_{n}\hat{b}^{\dagger}_{p}\hat{b}^{\dagger}_{q}\rangle\nonumber\\
&-i\hbar \mathcal{F}_{r}^{*}d^{{ln}^{*}} f^{n}\langle \hat{b}^{\dagger}_{r}\hat{b}^{\dagger}_{p}\hat{b}^{\dagger}_{q}\rangle\nonumber\\
&+i\hbar \mathcal{F}_{r}^{*}d^{{ln}^{*}} f^{l}\langle \hat{b}^{\dagger}_{r}\hat{b}^{\dagger}_{p}\hat{b}^{\dagger}_{q}\rangle\nonumber\\
&-i\gamma_{11}\langle \hat{c}^{\dagger}_{l}\hat{c}_{n}\hat{b}^{\dagger}_{p}\hat{b}^{\dagger}_{q}\rangle \label{24},
\end{align}
where $ f^{n} $ is the density of state, $ n $ and $\gamma_{11} $ act as dephasing source with elements defined in appendix A. The other spontaneous terms, $ \langle \hat{c}^{\dagger}_{n}\hat{c}_{m}\hat{b}^{\dagger}_{r}\hat{b}^{\dagger}_{p}\rangle $ and $ \langle \hat{c}^{\dagger}_{m}\hat{c}_{g}\hat{b}^{\dagger}_{q}\hat{b}^{\dagger}_{r}\rangle $ in Eqs. \eqref{22}  and \eqref{23}, are structurally similar to Eq. \eqref{24}
\begin{align}
i\hbar\dfrac{\partial}{\partial t}\langle \hat{c}^{\dagger}_{n}\hat{c}_{m}\hat{b}^{\dagger}_{r}\hat{b}^{\dagger}_{p}\rangle =&(-\varepsilon^{n}+\varepsilon^{m}-\hbar\omega_{p}-\hbar\omega_{r})\langle \hat{c}^{\dagger}_{n}\hat{c}_{m}\hat{b}^{\dagger}_{r}\hat{b}^{\dagger}_{p}\rangle\nonumber\\
&-i\hbar \mathcal{F}_{q}^{*}d^{{nm}^{*}}f^{m}\langle \hat{b}^{\dagger}_{q}\hat{b}^{\dagger}_{r}\hat{b}^{\dagger}_{p}\rangle\nonumber\\
&+i\hbar \mathcal{F}_{q}^{*}d^{{nm}^{*}}f^{n}\langle \hat{b}^{\dagger}_{q}\hat{b}^{\dagger}_{r}\hat{b}^{\dagger}_{p}\rangle\nonumber\\
&-i\gamma_{12}\langle \hat{c}^{\dagger}_{n}\hat{c}_{m}\hat{b}^{\dagger}_{r}\hat{b}^{\dagger}_{p}\rangle \label{25}.
\end{align}
and
\begin{align}
i\hbar\dfrac{\partial}{\partial t}\langle \hat{c}^{\dagger}_{m}\hat{c}_{g}\hat{b}^{\dagger}_{q}\hat{b}^{\dagger}_{r}\rangle =&(-\varepsilon^{m}+\varepsilon^{g}-\hbar\omega_{q}-\hbar\omega_{r})\langle \hat{c}^{\dagger}_{m}\hat{c}_{g}\hat{b}^{\dagger}_{q}\hat{b}^{\dagger}_{r}\rangle\nonumber\\
&+i\hbar \mathcal{F}_{p}^{*}d^{{mg}^{*}} f^{m}\langle \hat{b}^{\dagger}_{p}\hat{b}^{\dagger}_{q}\hat{b}^{\dagger}_{r}\rangle\nonumber\\
&-i\hbar \mathcal{F}_{q}^{*}d^{{mg}^{*}} f^{g}\langle \hat{b}^{\dagger}_{q}\hat{b}^{\dagger}_{r}\hat{b}^{\dagger}_{p}\rangle\nonumber\\
&-i\gamma_{22}\langle \hat{c}^{\dagger}_{m}\hat{c}_{g}\hat{b}^{\dagger}_{q}\hat{b}^{\dagger}_{r}\rangle \label{26}.
\end{align}
The explicit forms of $ \gamma_{12}$ and $\gamma_{22} $ are  given in appendix A.
 \section{analytic solution}\label{sec6}
The expectation value for any operator $ \hat{O} $ can be written according to the density matrix$ \rho $: 
\begin{equation}
\langle \hat{O}\rangle = Tr(\rho \hat{O}).\label{a1}
\end{equation} 
For our subsequent calculations it is advantageous to make the unitary transformation $ \overline{\rho}=U^{-1}\rho U $, in it,
\begin{align}
U(t)=e^{-iH_{0}t/\hbar},\label{a2}
\end{align}
where $ H_{0} $ has been given in  Eq.\eqref{20}. Substituting $ \rho $ by $ \overline{\rho} $ in Eq. \eqref{a1}, the expectation value yields,
\begin{align}
Tr(\rho\hat{O})=Tr(U\overline{\rho} U^{-1}\hat{O})=Tr(e^{-iH_{0}t/\hbar}\overline{\rho} e^{iH_{0}t/\hbar}\hat{O}).\label{a3}
\end{align}
 Considering the general property of the trace $ Tr(ABC)=Tr(CAB) $,
\begin{equation}
Tr(e^{-iH_{0}t/\hbar}\overline{\rho} e^{iH_{0}t/\hbar}\hat{O})=Tr(\overline{\rho} e^{iH_{0}t/\hbar}\hat{O}e^{-iH_{0}t/\hbar}).\label{a4}
\end{equation}
Accordingly, for  the operators appeared in the Eqs. \eqref{21}-\eqref{26}, by using the following relation,
\begin{equation*}
e^{-\alpha \hat{A}}\hat{O}e^{\alpha \hat{A}}=\hat{O}-\alpha[\hat{A},\hat{O}]+\dfrac{\alpha^{2}}{2!}[\hat{A},[\hat{A},\hat{O}]]+...
\end{equation*}
the expectation values can be calculated, 
\begin{align}
&\langle \hat{c}^{\dagger}_{l}\hat{c}_{n}\hat{b}^{\dagger}_{p}\hat{b}^{\dagger}_{q}\rangle =\overline{\langle \hat{c}^{\dagger}_{l}\hat{c}_{n}\hat{b}^{\dagger}_{p}\hat{b}^{\dagger}_{q}\rangle} e^{i(\varepsilon^{l}-\varepsilon^{n}+\hbar\omega_{p}+\hbar\omega_{q})t/\hbar}\nonumber\\
&\langle \hat{b}^{\dagger}_{p}\hat{b}^{\dagger}_{q}\hat{b}^{\dagger}_{r}\rangle =\overline{\langle \hat{b}^{\dagger}_{p}\hat{b}^{\dagger}_{q}\hat{b}^{\dagger}_{r}\rangle} e^{i(\hbar\omega_{r}+\hbar\omega_{p}+\hbar\omega_{q})t/\hbar},\label{a5}
\end{align}
where
\begin{align}
&\overline{\langle \hat{c}^{\dagger}_{l}\hat{c}_{n}\hat{b}^{\dagger}_{p}\hat{b}^{\dagger}_{q}\rangle}=\langle\overline{\rho} (\hat{c}^{\dagger}_{l}\hat{c}_{n}\hat{b}^{\dagger}_{p}\hat{b}^{\dagger}_{q})\rangle\nonumber\\
&\overline{\langle \hat{b}^{\dagger}_{p}\hat{b}^{\dagger}_{q}\hat{b}^{\dagger}_{r}\rangle}=\langle\overline{\rho} (\hat{b}^{\dagger}_{p}\hat{b}^{\dagger}_{q}\hat{b}^{\dagger}_{r})\rangle. \label{a6}
\end{align}
By inserting Eqs. \eqref{a5} in Eq. \eqref{24},  and considering the decay term as a exponential term, then  differentiation  with respect to the t, the solution will be given by:
\begin{align}
\overline{\langle \hat{c}^{\dagger}_{l}\hat{c}_{n}\hat{b}^{\dagger}_{p}\hat{b}^{\dagger}_{q}\rangle} =&\dfrac{i\hbar \mathcal{F}_{r}^{*}d^{{ln}^{*}}(f^{l}-f^{n})\overline{\langle \hat{b}^{\dagger}_{p}\hat{b}^{\dagger}_{q}\hat{b}^{\dagger}_{r}\rangle}}{i\hbar(\dfrac{i}{\hbar}(-\varepsilon^{l}+\varepsilon^{n}+\hbar\omega_{r}-i\hbar\gamma_{11}))}\nonumber\\
&[\exp(\dfrac{i}{\hbar}(-\varepsilon^{l}+\varepsilon^{n}+\hbar\omega_{r}-i\hbar\gamma_{11})t)-1].\label{27}
\end{align}
And so on for Eqs. \eqref{25} and \eqref{26} we have,
\begin{align}
\overline{\langle \hat{c}^{\dagger}_{n}\hat{c}_{m}\hat{b}^{\dagger}_{r}\hat{b}^{\dagger}_{p}\rangle} =&\dfrac{i\hbar \mathcal{F}_{q}^{*}d^{{nm}^{*}}(f^{n}-f^{m})\overline{\langle \hat{b}^{\dagger}_{p}\hat{b}^{\dagger}_{q}\hat{b}^{\dagger}_{r}\rangle}}{i\hbar(\dfrac{i}{\hbar}(-\varepsilon^{n}+\varepsilon^{m}+\hbar\omega_{q}-i\hbar\gamma_{12}))}\nonumber\\
&[\exp(\dfrac{i}{\hbar}(-\varepsilon^{n}+\varepsilon^{m}+\hbar\omega_{q}-i\hbar\gamma_{12})t)-1].\label{a7}\\
\overline{\langle \hat{c}^{\dagger}_{m}\hat{c}_{g}\hat{b}^{\dagger}_{q}\hat{b}^{\dagger}_{r}\rangle} =&\dfrac{i\hbar \mathcal{F}_{p}^{*}d^{{mg}^{*}}(f^{m}-f^{g})\overline{\langle \hat{b}^{\dagger}_{p}\hat{b}^{\dagger}_{q}\hat{b}^{\dagger}_{r}\rangle}}{i\hbar(\dfrac{i}{\hbar}(-\varepsilon^{m}+\varepsilon^{g}+\hbar\omega_{p}-i\hbar\gamma_{22}))}\nonumber\\
&[\exp(\dfrac{i}{\hbar}(-\varepsilon^{m}+\varepsilon^{g}+\hbar\omega_{p}-i\hbar\gamma_{22})t)-1].\label{a8}
\end{align}
If we transform  the expectation value in Eqs. \eqref{22} and \eqref{23} according to the Eq.\eqref{a4}  and  inserting Eqs. \eqref{27}-\eqref{a8} into these equations, 
\begin{align}
&\overline{\langle\hat{c}^{\dagger}_{l}\hat{c}_{m}\hat{b}_{p}^{\dagger}\rangle} =\nonumber\\
&\lbrace\dfrac{i\hbar \mathcal{F}_{q}^{*}d^{{nm}^{*}}i\hbar \mathcal{F}_{r}^{*}d^{{ln}^{*}}(f^{l}-f^{n})\overline{\langle \hat{b}^{\dagger}_{p}\hat{b}^{\dagger}_{q}\hat{b}^{\dagger}_{r}\rangle}}{(-\varepsilon^{l}+\varepsilon^{n}+\hbar\omega_{r}-i\hbar\gamma_{11})(-\varepsilon^{l}+\varepsilon^{m}+\hbar\omega_{r}+\hbar\omega_{q}-i\hbar\gamma_{1})}\nonumber\\
&-\dfrac{i\hbar \mathcal{F}_{q}^{*}d^{{nm}^{*}}i\hbar \mathcal{F}_{r}^{*}d^{{ln}^{*}}(f^{n}-f^{m})\overline{\langle \hat{b}^{\dagger}_{p}\hat{b}^{\dagger}_{q}\hat{b}^{\dagger}_{r}\rangle}}{(-\varepsilon^{n}+\varepsilon^{m}+\hbar\omega_{q}-i\hbar\gamma_{12})(-\varepsilon^{l}+\varepsilon^{m}+\hbar\omega_{r}+\hbar\omega_{q}-i\hbar\gamma_{1})}\rbrace\nonumber\\
&(\exp(\dfrac{i}{\hbar}(-\varepsilon^{l}+\varepsilon^{m}+\hbar\omega_{r}+\hbar\omega_{q}-i\hbar\gamma_{1})t)-1)\label{28}.
\end{align}
and
\begin{align}
&\overline{\langle\hat{c}^{\dagger}_{n}\hat{c}_{g}\hat{b}_{r}^{\dagger}\rangle} =\nonumber\\
&\lbrace\dfrac{i\hbar \mathcal{F}_{q}^{*}d^{{nm}^{*}}i\hbar \mathcal{F}_{p}^{*}d^{{mg}^{*}}(f^{n}-f^{m})\overline{\langle \hat{b}^{\dagger}_{p}\hat{b}^{\dagger}_{q}\hat{b}^{\dagger}_{r}\rangle}}{(-\varepsilon^{n}+\varepsilon^{m}+\hbar\omega_{q}-i\hbar\gamma_{12})(-\varepsilon^{n}+\varepsilon^{g}+\hbar\omega_{p}+\hbar\omega_{q}-i\hbar\gamma_{2})}\nonumber\\
&-\dfrac{i\hbar \mathcal{F}_{q}^{*}d^{{nm}^{*}}i\hbar \mathcal{F}_{p}^{*}d^{{mg}^{*}}(f^{m}-f^{g})\overline{\langle \hat{b}^{\dagger}_{p}\hat{b}^{\dagger}_{q}\hat{b}^{\dagger}_{r}\rangle}}{(-\varepsilon^{m}+\varepsilon^{g}+\hbar\omega_{p}-i\hbar\gamma_{22})(-\varepsilon^{n}+\varepsilon^{g}+\hbar\omega_{p}+\hbar\omega_{q}-i\hbar\gamma_{2})}\rbrace\nonumber\\
&(\exp(\dfrac{i}{\hbar}(-\varepsilon^{n}+\varepsilon^{g}+\hbar\omega_{p}+\hbar\omega_{q}-i\hbar\gamma_{2})t)-1)\label{29}.
\end{align}
By replacing these results into Eq. \eqref{21}, the final form can be obtained. Now, we consider some assumptions for description the three photon absorption process. The states $n$ and $m$ are the virtual states, so their  populations are vanishes. In the three photon absorption, in Fig \ref{pz4}, we assume  the  involved frequencies   are the same, so that,   
$ \omega_{p}=\omega_{q}=\omega_{r}=\omega $ so ${F}_{p}={F}_{q}={F}_{r}={F}  $ and $\hat{b}^{\dagger}_{q}=\hat{b}^{\dagger}_{p}=\hat{b}^{\dagger}_{r}=\hat{b}^{\dagger} $. On the other hand the virtuale states are nearly coincident with one of the resonant real level. Therefore we consider $ \hbar\omega^{ln}=\hbar\omega^{nm}=\hbar\omega^{mg} $ .  For simplicity the decay rate of states has been considered as the same to each other as $ \gamma $. According to this  assumption, 
\begin{align}
\overline{\langle \hat{c}_{l}^{\dagger}\hat{c}_{g}\rangle} =&\nonumber\\
&\dfrac{i\hbar \mathcal{F}_{r}^{*}d^{{ln}^{*}}i\hbar \mathcal{F}_{q}^{*}d^{{nm}^{*}}i\hbar \mathcal{F}_{p}^{*}d^{{mg}^{*}}(f^{l}-f^{g})\overline{\langle \hat{b}^{\dagger}_{p}\hat{b}^{\dagger}_{q}\hat{b}^{\dagger}_{r}\rangle}}{\hbar^{3}(\omega^{ln}-\omega +i\gamma)(\omega^{lm}-2\omega +i\gamma)}\nonumber\\
&\dfrac{\exp(-i(\omega^{lg}-3\omega+i\gamma)t)-1}{(\omega^{lg}-3\omega +i\gamma)}\label{30},
\end{align}
According to Eqs.\eqref{a4} and \eqref{a5} one obtains,
 \begin{align}
&\langle  \hat{c}_{l}^{\dagger}\hat{c}_{g}\rangle =\overline{\langle \hat{c}_{l}^{\dagger}\hat{c}_{g}\rangle}  e^{i(\varepsilon^{l}-\varepsilon^{g})t/\hbar}\nonumber\\
&\langle \hat{b}^{\dagger}_{p}\hat{b}^{\dagger}_{q}\hat{b}^{\dagger}_{r}\rangle =\overline{\langle \hat{b}^{\dagger}_{p}\hat{b}^{\dagger}_{q}\hat{b}^{\dagger}_{r}\rangle
} e^{i(\hbar\omega_{r}+\hbar\omega_{p}+\hbar\omega_{q})t/\hbar}\label{a8}
\end{align}
Inserting the above equations into the Eq. \eqref{30}, the suitable form for expectation value will be achieved, 
\begin{align}
\langle \hat{c}_{l}^{\dagger}\hat{c}_{g}\rangle =&\nonumber\\
&\dfrac{i\hbar \mathcal{F}_{r}^{*}d^{{ln}^{*}}i\hbar \mathcal{F}_{q}^{*}d^{{nm}^{*}}i\hbar \mathcal{F}_{p}^{*}d^{{mg}^{*}}(f^{l}-f^{g})\langle \hat{b}^{\dagger}_{p}\hat{b}^{\dagger}_{q}\hat{b}^{\dagger}_{r}\rangle}{\hbar^{3}(\omega^{ln}-\omega +i\gamma)(\omega^{lm}-2\omega +i\gamma)}\nonumber\\
&\dfrac{\exp(-i(\omega^{lg}-3\omega+i\gamma)t)-1}{(\omega^{lg}-3\omega +i\gamma)}\times\nonumber\\ &e^{i(\varepsilon^{l}-\varepsilon^{g})t/\hbar} e^{-i(\hbar\omega_{r}+\hbar\omega_{p}+\hbar\omega_{q})t/\hbar},\label{a9},
\end{align}
here $ \hbar\omega^{lm}=\varepsilon^{l}-\varepsilon^{m} $. Using Eq. \eqref{10} and Eq. \eqref{18}, one obtains the rate of three photon absorption in quantum approach,
\begin{align}
R^{3}_{l}=&\vert \dfrac{(i\hbar \mathcal{F}^{*}d^{{ln}^{*}})(i\hbar \mathcal{F}^{*}d^{{nm}^{*}})(i\hbar \mathcal{F}^{*}d^{{mg}^{*}})(f^{l}-f^{g})\langle \hat{b}^{\dagger}\hat{b}^{\dagger}\hat{b}^{\dagger}\rangle}{\hbar^{3}(\omega^{ln}-\omega +i\gamma)(\omega^{lm}-2\omega +i\gamma)}\vert^{2}\nonumber\\
&\times 2\pi \rho(\omega^{lg}-3\omega)\label{31},
\end{align}
where $ \rho(\omega^{lg}-3\omega) $ is the density of final state and Lorentzian shape is a well-known example of it and its derivation is given in Appendix B. In quantum approach the electric field has the form as $ E=E^{-}+E^{+} $ and the negative part defined as  $E^{-}\equiv{i\hbar \mathcal{F}^{*} \hat{b}^{\dagger}} $.  Therefore the form of the absorption rate  in quantum and semi-classical approaches is the same (see the Eq. (\ref{11})). 
By considering Eq. \eqref{h21} one can obtain the expectation value $ \langle \hat{b}^{\dagger}\hat{b}^{\dagger}\hat{b}^{\dagger}\rangle $ for coherent and squeezed state.
The normalized rate of the third order absorption for squeezed and coherent lights with the same intensity, are illustrated in Fig. \ref{p5}, where $ \gamma=2\pi(1\times 10^{13})\dfrac{Rad}{Sec} $, $d^{ln}=d^{nm}=d^{mg}= 8\times 10^{-30}   $cm$,  \vert \alpha\vert^{2}=7 $ and different values of the squeezed parameter have been chosen.  
\begin{figure}[ht]
\begin{center}
\includegraphics[scale=0.35]{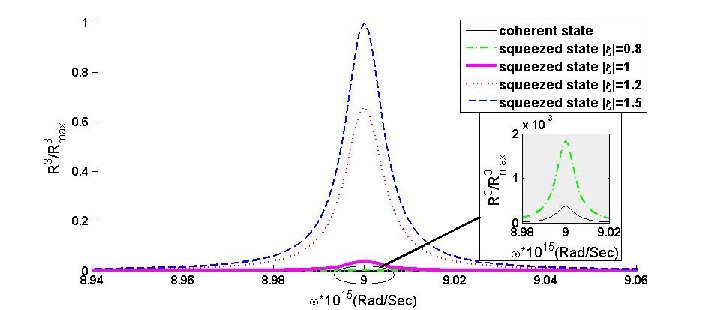}\label{p5}
\end{center}
\caption{The rate of three photon absorption for squeezed  and coherent light with the same intensity and various squeezed parameters ($\vert\xi\vert$ = 0.8,  $\vert\xi\vert$ = 1,  $\vert\xi\vert$ = 1.2  and $\vert\xi\vert$ = 1.5 , $ \varphi_{\xi}=0 $. $ R^{3}_{max} $ is related to the maximum value of squeezed light with $\vert\xi\vert$ = 1.5  ).}\label{p5}
\end{figure}

This figure shows that nonlinear media are very sensitive to the state of light. When the incident light is in a  quantum state, specially squeezed one, the rate of absorption is several times greater  than the case if the  coherent light is applied. On the other hand, one can see obviously that  for the same intensity, by increasing the squeezed parameter the difference of the rate absorption between coherent and squeezed light becomes very prominent. 

This important point is an indication for the accepted result, that the nonlinearity can be used for amplification of non-classicality. So that the proposed system can be considered as a detector for  the squeezed state of radiation. 

Investigating figures \ref{plot1} and \ref{p5} shows that the photon generation by a driving field in squeezed state, in 1D cavity with moving wall and three photon absorption in nonlinear medium, is several times greater than the case of coherent light. It seems that this phenomenon may  be explained by considering that the noises in squeezed light, at least in one of its quadrature, is suppressed, and altogether  the squeezed light shows more interaction power in nonlinear media.\\
 
\section{conclusion}\label{sec7}
The main purposes of the paper are twofold. First, we have studied one-dimensional cavity with moving mirror in presence of squeezed  and coherent light. The photon creation rate has shown  an obvious difference between these two modes of light. The photon generation rate increased with increasing of squeezing parameter. 

Second, by another approach, we have demonstrated that the nonlinear responses of the media is highly affected by the state of the interaction field. We have illustrated   the third order of the photon absorption rate of squeezed light in comparison to  the coherent  case, dependent on the squeezing parameter, indicates drastic difference. 

As most important result, we conclude that the nonlinear effects in 1D cavity with moving mirror are comparable with a nonlinear medium, when they interact with different states of light. In other words, the nonlinearity amplifies the non-classicality of states.  We have seen the drastic difference between squeezed and coherent states of light, which can be used for detection the non-classicality of states. 
\appendix
\section{}
 The decaying terms in Eqs. \eqref{27}-\eqref{29} have the following forms,
\begin{align}
i\gamma_{11}=
&-i\hbar \mathcal{F}_{q}d^{nm}\langle \hat{c}^{\dagger}_{l}\hat{c}_{m}\hat{b}_{q}\hat{b}^{\dagger}_{p}\hat{b}^{\dagger}_{q}\rangle\nonumber\\
&-i\hbar \mathcal{F}_{p}d^{mg}\langle\hat{c}^{\dagger}_{m}\hat{c}_{g}\hat{c}^{\dagger}_{l}\hat{c}_{n}
\hat{b}^{\dagger}_{q}\rangle\nonumber\\
&-i\hbar \mathcal{F}_{q}d^{nm}\langle\hat{c}^{\dagger}_{n}\hat{c}_{m}\hat{c}^{\dagger}_{l}
\hat{c}_{n}\hat{b}^{\dagger}_{p}\rangle,
\end{align}

\begin{align}
i\gamma_{12}=
&-i\hbar \mathcal{F}_{r}d^{ln}\langle \hat{c}^{\dagger}_{l}\hat{c}_{m}\hat{b}_{r}\hat{b}^{\dagger}_{p}\hat{b}^{\dagger}_{r}\rangle\nonumber\\
&+i\hbar \mathcal{F}_{p}d^{mg}\langle \hat{c}^{\dagger}_{n}\hat{c}_{g}\hat{b}_{p}\hat{b}^{\dagger}_{p}\hat{b}^{\dagger}_{r}\rangle\nonumber\\
&-i\hbar \mathcal{F}_{r}d^{ln}\langle\hat{c}^{\dagger}_{l}\hat{c}_{n}\hat{c}^{\dagger}_{n}\hat{c}_{m}
\hat{b}^{\dagger}_{p}\rangle\nonumber\\
&-i\hbar \mathcal{F}_{p}d^{mg}\langle\hat{c}^{\dagger}_{m}\hat{c}_{g}\hat{c}^{\dagger}_{n}
\hat{c}_{m}\hat{b}^{\dagger}_{r}\rangle, 
\end{align}
and 
\begin{align}
i\gamma_{22}=
&-i\hbar \mathcal{F}_{q}d^{nm}\langle \hat{c}^{\dagger}_{n}\hat{c}_{g}\hat{b}_{q}\hat{b}^{\dagger}_{q}\hat{b}^{\dagger}_{r}\rangle\nonumber\\
&-i\hbar \mathcal{F}_{q}d^{nm}\langle\hat{c}^{\dagger}_{n}\hat{c}_{m}\hat{c}^{\dagger}_{m}\hat{c}_{g}
\hat{b}^{\dagger}_{r}\rangle\nonumber\\
&-i\hbar \mathcal{F}_{r}d^{ln}\langle\hat{c}^{\dagger}_{l}\hat{c}_{n}\hat{c}^{\dagger}_{m}
\hat{c}_{g}\hat{b}^{\dagger}_{q}\rangle.
\end{align}
\section{}
According to Ref. \cite{06}, the time dependence of  the square of  absolute value of  the expectation value can be obtained,
\begin{align}
\vert \dfrac {\exp(-i(\omega^{lg}-3\omega+i\gamma)t)-1}{(\omega^{lg}-3\omega +i\gamma)}\vert^{2}=\dfrac{4\sin^{2}((\omega^{lg}-3\omega +i\gamma)t/2)}{(\omega^{lg}-3\omega+i\gamma)^{2}}
\end{align}
On the other hand, for large values of the time, one can prove, 
\begin{equation}
\text{lim}_{t\rightarrow \infty} t^{2}\dfrac{\sin^{2}(x)}{x^{2}}=2\pi t\delta(\omega^{lg}-3\omega)\quad \text{for} \quad x=(\omega^{lg}-3\omega)t/2 .
\end{equation} 
In real situation, the final state is not sharp in the  frequency transition and make a  continuous distribution in  frequency domain. So the $ \delta(\omega^{lg}-3\omega) $ can be replaced by $ \rho(\omega^{lg}-3\omega) $. Usually, the $ \rho(\omega^{lg}-3\omega) $  has  Lorentzian shape including  the decay term $\gamma $.  So the absorption  probability can be written in the form, 
\begin{align*}
P^{3}_{l}=&\vert \dfrac{i\hbar \mathcal{F}^{*}d^{{ln}^{*}}i\hbar \mathcal{F}^{*}d^{{nm}^{*}}i\hbar \mathcal{F}^{*}d^{{mg}^{*}}(f^{l}-f^{g})\langle \hat{b}^{\dagger}\hat{b}^{\dagger}\hat{b}^{\dagger}\rangle}{\hbar^{3}(\omega^{ln}-\omega +i\gamma)(\omega^{lm}-2\omega +i\gamma)}\vert^{2}\nonumber\\
&\times 2\pi t \rho(\omega^{lg}-3\omega).
\end{align*}
Using Eq.\eqref{10} one gets  the Eq.\eqref{31}.


\begin{thebibliography}{99}


\bibitem{Dodonov1}{V.V. Dodonov, "Resonance excitation and cooling of electromagnetic modes in a cavity with an oscillating wall," Phys. Lett .A \textbf{213}, 219-225 (1996).}  
\bibitem{Dodonov2}{V.V. Dodonov, "Photon creation and excitation of a detector in a cavity with a resonantly vibrating wall," Phys. Lett .A \textbf{207}, 126-132 (1995).} 
\bibitem{Dodonov3}{V.V. Dodonov, A.B. Klimov, D.E. Nikonov, "Quantum phenomena in non-stationary media," Phys. Lett .A \textbf{47}, 4422 (1993).} 

\bibitem{Yablonovitch}{E. Yablonovitch, "Accelerating reference frame for electromagnetic waves in a rapidly growing plasma: Unruh-Davies-Fulling-DeWitt radiation and the non-adiabatic Casimir effect," Phys. Rev. Lett \textbf{62}, 15 (1989).} 
\bibitem{Schwinger}{J. Schwinger, "Casimir energy for dielectrics," Proc. Natl. Acad. Sci. USA, \textbf{89}, 4091-4093 (1992).}
\bibitem{Maia}{P. A. Maia Neto, L. A. S. Machado, "Quantum radiation generated by a moving mirror in free space," Phys. Rev. A \textbf{54}, 3420 (1996).}
\bibitem{Dodonov4}{V.V. Dodonov, A.B. Klimov, "Generation and detection of photons in a cavity with a resonantly oscillating boundary," Phys. Rev. A \textbf{53}, 2665 (1996).} 
\bibitem{Wilson}{C. M. Wilson, G. Johansson, A. Pourkabirian, M. Simoen, J. R. Johansson, T. Duty, F. Nori and P. Delsing, "Observation of the dynamical Casimir effect in a superconducting circuit," Nature. V, \textbf{479}, 376 (2011).}
\bibitem{Dezael}{F. X. Dezael, A. Lambrecht, "Analogue Casimir radiation using an optical parametric oscillator," EPL \textbf{89}, 14001 (2010).}
\bibitem{Dodonov5}{V.V. Dodonov, "Nonstationary Casimir effect  and analytical solutions for quantum fields in cavities with moving boundaries," Arxiv: quant-ph/0106081V1 (2001).}
 \bibitem{Castro}{A. S. M. de Castro, A. Cacheffo, V. V. Dodonov, "Influence of the field-detector coupling strength on the dynamical Casimir effect," Phys. Rev. A \textbf{87}, 033809 (2013).}
 \bibitem{Dodonov7}{A. V. Dodonov, "Continuous intracavity monitoring of the dynamical Casimir effect," Phys. Scr.   \textbf{87}, 038103 (2013).}
 
\bibitem{Dodonov6}{V.V. Dodonov, A.B. Klimov, V. I. Manko, "Generation of squeezed states in a resonator with a moving wall," Phys. Lett .A \textbf{149}, 4 (1990).} 

\bibitem{Sousa}{I. M. de. Sousa, A. V. Dodonov, "Microscopic toy model for Cavity dynamical Casimir effet ," arXiv:1504.02413v1 [quant-ph] (2015).} 

\bibitem{Gong}{Z.R. Gong, H. Ian, Yu-Xi. Liu, C.P. Sun, and F. Nori, "Effective Hamiltonian approach to the Kerr nonlinearity in an optomechanical system," Phys. Rev. A \textbf{80}, 065801 (2009).}
\bibitem{Shapiro}{J. H. Shapiro, "Quantum noise and excess noise in optical homodyne and heterodyne receivers," IEEE J. Quant. Electron \textbf{QE-21}, 237-250 (1985).}
\bibitem{Teich}{M. C. Teich, B. E. A. Saleh, "Squeezed states of light ," Quantum Opt. l, 152-191 (1989).}
\bibitem{Breitenbach}{G. Breitenbach, S. Schiller, J. Mylnek, "Measurement of the quantum states of squeezed light," Nature, \textbf{387}, 471 - 475 (1997).}
\bibitem{Loundon}{R. Loudon, P. L. Knight, "Squeezed light ," J. Mod. Op, \textbf{34}, 709-759 (1987).}

\bibitem{Dalton}{B. J. Dalton, Z. Ficek, S. Swain, "Topical review atoms in squeezed light fields," J. Mod. Op, \textbf{46}, 379-474 (1999).}

\bibitem{Cessa}{H. Moya-Cessa, A. Vidiella. Barranco , "Interaction of squeezed light with two-level atoms ," J. Mod. Op, \textbf{39}, 2481-2499 (1992).}

\bibitem{02}
T. Tanimura, D. Akamatsu, Y. Yokoi, A. Furusawa, and M. Kozuma, "Generation of a squeezed vacuum resonant on a rubidium D1 line with periodically poled KTiOPO4," Opt. Lett. {\bf 31}, 2344 (2006).
\bibitem{03}
T. Aoki, G. Takahashi and A. Furusawa, "Squeezing at 946nm with periodically poled KTiOPO4," Opt. Express. {\bf 14}, 6930 (2006).
\bibitem{Zubairy}
M. S. Zubairy, M. S. K. Razmi, Saleem. Iqbal and M. Idress, "Squeezed states in a multiphoton absorption process," Phys. Lett .A {\bf 98}, 168 (1983).

\bibitem{Fabre}{C. Fabre, M. Pinard, S. Bourzeix, A. Heidmann, E. Giacobino, and S. Reynaud, "Quantum-noise reduction using a cavity with a movable mirror," Phys. Rev. A \textbf{49}, 1337 (1994).}
\bibitem{Mancini}{S. Mancini, V. I. Manko, P. Tombesi, "Ponderomotive control of quantum macroscopic coherence," Phys. Rev. A \textbf{55}, 3042 (1997).}
\bibitem{04}
G. Masada, T. Suzudo, Y. Satoh, H. Ishizuki, T. Taira, and A. Furusawa,  "Efficient generation of highly squeezed light with periodically poled MgO: LiNbO3," Opt. Express. {\bf 18}, 13114 (2010).
\bibitem{08}
M. Mehmet, H. Vahlbruch, N. Lastzka, K. Danzmann, and R. Schnabel, "Observation of squeezed states with strong photon-number oscillations,"  Phys. Rev. A {\bf 81}, 013814 (2010).
\bibitem{09}
E. S. Polzik, J. Carri and H. J. Kimble, "Atomic spectroscopy with squeezed light for sensitivity beyond the vacuum-state limit," Appl. Phys. B {\bf 55}, 279 (1992).
\bibitem{01}
H. Paul, U. Mohr and W. Brunner, "Change of photon statistics due to multi-photon absorption," Opt. Commun. {\bf 17}, 145 (1976).
\bibitem{05}
J. Gea-Banacloche, "Two-photon absorption of nonclassical light," Phys. Rev. Lett. {\bf 62}, 1603 (1989).
\bibitem{Enaki}{N. Enaki, M. Macovei, D. Mihalache, "Cooperative two-photon interaction with nonclassical light," Physica A \textbf{258}, 383-394 (1998).}

\bibitem{Schumaker}{B. L. Schumaker, C. M, Caves, "New formalism for two-photon quantum optics," Phys. Rev. A \textbf{31}, 3093 (1985).}
\bibitem{06}
R. W. Boyd, \emph{Nonlinear Optics} (Academic, Boston, 1992). 
\bibitem{07}
M. Kira and S. W. Koch, "Quantum-optical spectroscopy of semiconductors," Phys. Rev. A {\bf 73}, 013813 (2006).
\bibitem{010}
H. Haug, \emph{Quantum Theory of the Optical and Electronic Properties of Semiconductors} (Word Scientific, Singapore, 1994)


\bibitem{Weiss}U. Weiss, \emph{Quantum Dissipative Systems} (World Scientific, Singapore, 1993)
\bibitem{013}
L. Schneebeli,  M. Kira and W. Koch, "Characterization of strong light-matter coupling in semiconductor quantum-dot micro-cavities via photon-statistics spectroscopy," Phys. Rev. Lett. {\bf 101}, 097401 (2008).
\bibitem{014}
M. O. Scully and M. S. Zubairy, \emph{Quantum Optics} (Cambridge University Press, Cambridge, Engeland, 1997)
\bibitem{015}%
W. Vogel and D. G. Welsch, \emph{Quantum Optics An Introduction}(Wiley-VCH, Berlin, 2001)
\bibitem{Pinard}{M. Pinard, C. Fabre, and A. Heidmann, "Quantum-nondemolition measurement of light by a piezoelectric crystal," Phys. Rev. A \textbf{51}, 2443 (1995).}
\bibitem{Lakowicz}{J.R.Lakowicz, , \emph{Topics in Fluorescence Spectroscopy vol.5 Nonlinear and Two-photon Induced Fluorescence} (Kluwer Academic Publishers, 2002).}
\end{thebibliography}
\end{document}